\documentclass[iop]{emulateapj}

\newcommand\5{{\footnotesize V}}
\newcommand\4{{\footnotesize IV}}
\newcommand\3{{\footnotesize III}}
\newcommand\2{{\footnotesize II}}
\newcommand\1{{\footnotesize I}}
\newcommand\lam{{$\lambda$}}
\newcommand\vsini{{$v$\,sin$i$}}
\newcommand\kms{$\rm{km\,s^{-1}}$}

\newcommand\ph{\phantom{1}}

\shorttitle{Sk\,183: An O3-type star in the Wing of the SMC}
\shortauthors{Evans et al.}

\begin{document}


\title{A rare early-type star revealed in the Wing of the Small Magellanic Cloud}

\slugcomment{Received; Accepted}


\author{C. J. Evans\altaffilmark{1}, 
R. Hainich\altaffilmark{2},
L. M. Oskinova\altaffilmark{2},
J. S. Gallagher III\altaffilmark{3},
Y.-H. Chu\altaffilmark{4}, \\
R. A. Gruendl\altaffilmark{4},
W.-R. Hamann\altaffilmark{2},
V. H\'{e}nault-Brunet\altaffilmark{5}
{\sc and} H. Todt\altaffilmark{2}
}

\altaffiltext{1}{UK Astronomy Technology Centre, 
                 Royal Observatory Edinburgh, 
                 Blackford Hill, 
                 Edinburgh, 
                 EH9 3HJ, UK}
\altaffiltext{2}{Institute for Physics and Astronomy, 
                 University of Potsdam, 
                 14476 Potsdam, 
                 Germany}
\altaffiltext{3}{Department of Astronomy, University of Wisconsin-Madison, 
                 5534 Sterling, 475 North Charter St., 
                 Madison, WI 53706, USA}
\altaffiltext{4}{Astronomy Department, University of Illinois, 
                 1002 W. Green Street, Urbana, IL 61801, USA}
\altaffiltext{5}{Scottish Universities Physics Alliance (SUPA),
                 Institute for Astronomy, 
                 University of Edinburgh, 
                 Royal Observatory Edinburgh, 
                 Blackford Hill, 
                 Edinburgh, 
                 EH9 3HJ, UK}

\begin{abstract}
  Sk\,183 is the visually-brightest star in the N90 nebula, a
    young star-forming region in the Wing of the Small Magellanic
  Cloud (SMC). We present new optical spectroscopy from the Very Large
  Telescope which reveals Sk\,183 to be one of the most massive O-type
  stars in the SMC. Classified as an O3-type dwarf on the basis of its
  nitrogen spectrum, the star also displays broadened He~\1 absorption
  which suggests a later type. We propose that Sk\,183 has a composite
  spectrum and that it is similar to another star in the SMC, MPG~324.
  This brings the number of rare O2- and O3-type stars known in the
  whole of the SMC to a mere four. We estimate physical parameters
  for Sk\,183 from analysis of its spectrum. For a single-star model,
  we estimate an effective temperature of 46$\pm$2\,kK, a low
  mass-loss rate of $\sim$\,10$^{-7}$\,$M_\odot$\,yr$^{-1}$, and a
  spectroscopic mass of 46$^{+9}_{-8}$\,$M_\odot$ (for an adopted
    distance modulus of 18.7\,mag to the young population in the SMC
    Wing). An illustrative binary model requires a slightly hotter
  temperature ($\sim$47.5\,kK) for the primary component. In either
  scenario, Sk\,183 is the earliest-type star known in N90 and will
  therefore be the dominant source of hydrogen-ionising photons. This
  suggests Sk\,183 is the primary influence on the star formation along the
  inner edge of the nebula.
\end{abstract}

\keywords{open clusters and associations: individual (N90; NGC\,602) ---
stars: early-type --- stars: fundamental parameters --- stars: individual
(Sanduleak 183)}

\section{Introduction}\label{intro}

The Magellanic Clouds provide an excellent example of interacting,
metal-poor dwarf galaxies on our own doorstep, with neutral hydrogen
maps revealing large structures such as the Magellanic Bridge (which
connects the Clouds), and the extensive Magellanic Stream and Leading
Arm (e.g., Putman 2000). The Bridge is linked to the eastern side of
the Small Magellanic Cloud (SMC) by an extended `Wing', which has a
lower content of gas, dust and stars than the main body of the SMC,
suggesting it should be unfavourable to star formation. However, the
young stellar cluster NGC\,602 presents spectacular evidence to the
contrary, with a star-formation rate which appears comparable to
well-studied Galactic regions (e.g. Cignoni et al.  2009).

NGC\,602 is a loose term for a rich region comprising at least three
distinct clusters (Westerlund, 1964). NGC\,602a\footnote{Cluster 105
  from Lindsay (1958)} is embedded in the N90 emission nebula (Henize,
1956), NGC\,602b is a closely-related, smaller cluster approximately
1\farcm5 to the north, and NGC\,602c is $\sim$11$'$ to the northeast.
Other nearby H$\alpha$ emission regions include N88 and N89
(Lindsay~104) which, together with NGC\,602, are associated with
the only H$\alpha$ supergiant shell known in the SMC (`SGS-SMC1',
Meaburn, 1980).

Optical imaging of NGC\,602a from the {\em Hubble Space Telescope
  (HST)} Advanced Camera for Surveys (ACS) is particularly
striking\footnote{http://hubblesite.org/newscenter/archive/releases/2007/04/},
engendering the popular moniker of the `Hubble Bubble'. The {\em HST}
observations and deep infrared images from the {\em Spitzer Space
  Telescope} have given us an exquisite view of the stellar content
and star formation in NGC\,602a and N90 (Carlson et al. 2007, 2011;
Gouliermis et al. 2007; Schmalzl et al. 2008; Nigra et al. 2008;
Cignoni et al. 2009).  

\begin{figure}
\begin{center}
\includegraphics[width=8.5cm]{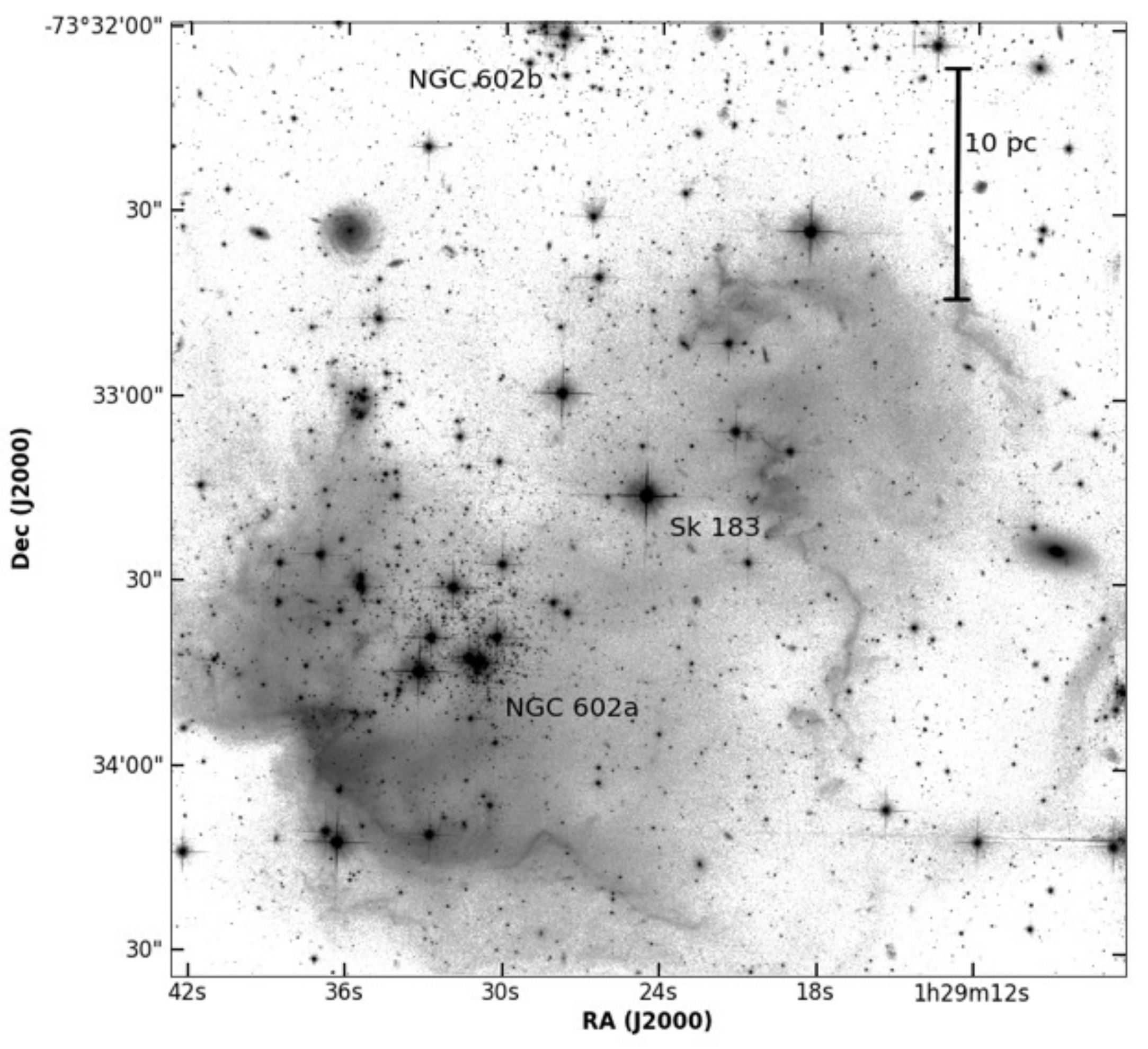}
\caption{{\em HST}--ACS {\it F814W} image of NGC\,602a and its associated H~\2
  region (N90), centred on Sk\,183. NGC\,602b is approximately 1\farcm5 to the 
north. The 10\,pc scale assumes the `short' distance modulus of 18.7\,mag (see
Section~\ref{reddening}).}\label{602}
\end{center}
\end{figure}

As part of a new study of SGS-SMC1 we have obtained optical
spectroscopy of its luminous stellar population with the Very Large
Telescope (VLT). In the course of the program we observed Sk\,183
(Sanduleak, 1969)\footnote{ $\alpha$\,=\,01$^{\rm h}$\,29$^{\rm
    m}$\,24\fs6, $\delta$\,=\,$-$73$^\circ$\,33$'$\,16\farcs43, J2000.
  Aliases include: NGC\,602a~\#15 (Westerlund, 1964); NGC\,602~\#8
  (Hutchings et al. 1991); SMC\,83235 (Massey, 2002).}, the visually
brightest star in the N90 nebula (see Figure~\ref{602}). In
this article we present the classification and analysis of these
observations, which reveal it to be one of the most massive O-type
stars in the SMC.

\section{Spectroscopy}

Optical spectroscopy was obtained on 2010 October 24-26 with the VLT
FLAMES instrument (Pasquini et al. 2002), as summarised in
Table~\ref{obsinfo}. Sk\,183 was observed as part of the larger survey
of the region with the Medusa-fibre mode of FLAMES, using three of the
standard settings of the Giraffe spectrograph: LR02, LR03, and HR15N,
delivering spectral resolving powers of 7000, 8500, and 16\,000,
respectively (e.g. Evans et al.  2011). Individual exposure times
were 1800s for all settings.

The ESO Common Pipeline Library FLAMES reduction routines (v.2.8.7)
were used for the standard processing stages, i.e. bias subtraction,
fibre location, summed extractions, division by a normalized
flat-field, and wavelength calibration. The spectra were then
corrected to the heliocentric frame, and a median sky spectrum was
subtracted (for more details see Evans et al. 2011).

Qualitative inspection of the spectra revealed no evidence of radial
velocity variations associated with binarity, further supported by the
measurements in Section~\ref{rv}. Thus, the spectra from each
exposure/setting were co-added, yielding a signal-to-noise ratio
in excess of 250 per rebinned pixel from the pipeline. However, note
that the observations only spanned three nights, so our detection
sensitivity to radial velocity variations from anything but a
short-period companion will have been low.

There are archival, low-dispersion observations of Sk\,183 from the
{\it International Ultraviolet Explorer (IUE)} from Hutchings et al.
(1991). Although at low resolution, the lack of strong UV
emission-lines is evident (see Figure~4 from Hutchings et al.),
consistent with the weak wind-profiles seen in other massive stars in
the SMC, e.g. Walborn et al. (1995, 2000).

\subsection{Spectral Classification}\label{class}

\begin{figure*}
\begin{center}
\includegraphics[angle=-90,width=17cm]{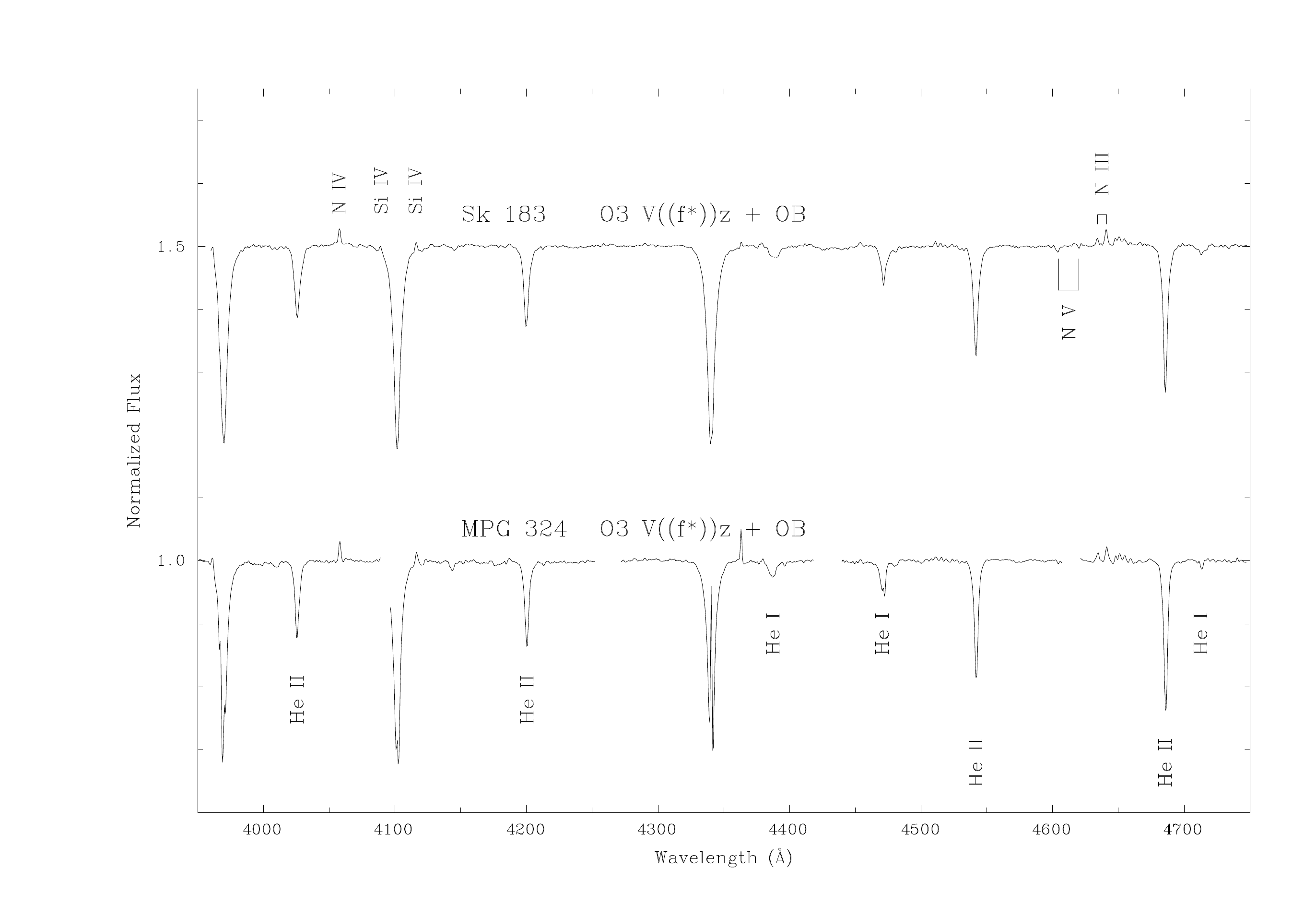}
\caption{Combined FLAMES spectrum of Sk\,183 together with that of
  NGC\,346 MPG~324 from Evans et al. (2006); both spectra have been
  smoothed and rebinned to an effective resolving power of 4000.
  Emission lines identified in Sk\,183 are N~\4 \lam4058; Si~\4
  \lam\lam4089, 4116; N~\3 \lam\lam4634-41-42, with absorption from
  N~\5 \lam\lam4604, 4620. Absorption lines identified in MPG~324 are
  He~\1 \lam\lam4388, 4471, 4713; He~\2 \lam\lam4026, 4200, 4542,
  4686. Both spectra show a combination of N~\4 emission and He~\1
  absorption inconsistent with a unique spectral type, suggestive of
  composite spectra. The gaps in the spectrum of MPG~324 are due to
  `hot' pixels in the detector.}\label{spectra}
\end{center}
\end{figure*}

The \lam\lam3950-4750\,\AA\/ region of the spectra of Sk\,183 is shown in
Figure~\ref{spectra}, in which the data have been smoothed and
rebinned to an equivalent resolving power of $R$\,$=$\,4000 for
classification (cf. the new Galactic atlas of standards from Sana et
al. in preparation). Although relatively modest in strength, N~\4
\lam4058 emission has an intensity comparable to that from N~\3
\lam\lam4634-41-42. Employing the criteria from Walborn et al. (2002), this
suggests a classification of O3~V((f$^\ast$))z, in which the `z'
suffix indicates that the \lam4686 absorption has a greater depth than
any of the other helium lines (Walborn \& Blades, 1997; Walborn 2009).

An early-type classification is supported by the weak
absorption from Nv (\lam\lam 4604, 4620), but there are other features
which are not fully consistent with the O3 classification. For
instance, the apparent intensities of the He~\1 \lam\lam 4388, 4471,
4713 lines suggest a later spectral type. The intensity of the He~\2
\lam4026 absorption is also relatively strong for an O3 type dwarf,
but may be explained by a contribution from He~\1. We note that
Hutchings et al.  (1991) classified their low-dispersion optical
spectrum of Sk\,183 as O6, but the lack of strong He~\1 \lam4471
absorption in their data argues for an earlier type.

\begin{center}
\begin{deluxetable}{lcccc}
\tabletypesize{\footnotesize}
\tabletypesize{\footnotesize}
\tablewidth{0pc}
\tablecolumns{5}
\tablecaption{Summary of VLT-FLAMES spectroscopy of Sk\,183\label{obsinfo}}
\tablehead{\colhead{Setting}  & \colhead{Date} & \colhead{Exp. Time} 
& \colhead{\lam-coverage} & $R$ \\
\colhead{} & \colhead{} & \colhead{[s]} & \colhead{[\AA]} & }
\startdata
LR02 & 2010-10-24 & 7$\times$1800 & 3960-4564 & $\phantom{1}$7000 \\
& 2010-10-26 & 2$\times$1800 & & \\
LR03 & 2010-10-25 & 4$\times$1800 & 4499-5071 & $\phantom{1}$8500 \\
HR15N & 2010-10-25 & 2$\times$1800 & 6442-6817 & 16000 
\enddata
\tablecomments{The final column gives the effective resolving power ($R$) at the
central wavelength of each setting.}
\end{deluxetable}
\end{center}

\vspace{-0.35in}
As a possible explanation of the spectral properties of Sk 183 we
propose that its spectrum is composite in nature, and results either
from a binary system or an unresolved multiple. While the star appears
as a single point source in the image from the {\em HST} ACS, the
spatial resolution in the Wide Field Channel (2~pixels\,$=$\,0\farcs1)
subtends several thousand AU at the distance of the SMC. The composite
character of the Sk 183 spectrum is further supported by the
morphology of the He~\1 absorption, which indicates rapid rotation in
a secondary component (e.g. \lam4388 in Figure~\ref{spectra}) or
perhaps evidence for further multiplicity.  Without additional data
(e.g., time-series photometry or multi-epoch spectroscopy) it is hard to
constrain the nature of Sk\,183 further and we adopt a classification
of O3~V((f*))z\,$+$\,OB. Similar composite classifications were
employed for three objects by Walborn et al. (2002), who demonstrated
that one (Cyg~OB2-22) was indeed a composite of an O3 supergiant and
an O6 dwarf.

Circumstantial evidence for the composite nature of Sk\,183 is also
provided by the spectra of another star in the SMC, MPG~324 (Massey et
al.  1989)\footnote{Aliases: NGC\,346~\#6 (Walborn et al.  1995);
  NGC\,346~007 (Evans et al.  2006).}, previously classified as an O4
dwarf (Niemela et al. 1986; Massey et al. 1989; Walborn et al.  1995,
2000; Evans et al. 2006). Returning to the FLAMES spectra of MPG~324
from Evans et al. (2006), similar features are seen as those in
Sk\,183, i.e., N~\4 emission indicative of an early ($<$O4) type,
combined with He~\1 absorption suggesting a later type (see
Figure~\ref{spectra}). Indeed, Evans et al. (2006) noted MPG~324 as a
single-lined binary due to a $\sim$30\,\kms\/ shift in the He~\2
\lam4542 line between two epochs of observations with one of the
FLAMES settings.  The He~\1 \lam4471 line in these two epochs was
contaminated by nebular emission, but the wings of the profile were
(qualitatively) consistent with a zero velocity shift, suggesting a
potentially different origin to the He~\2 absorption. We reclassify
MPG~324 as O3~V((f*))z\,$+$\,OB, and note that Massey et al. (2009)
classified MPG~324 as O5~V in the course of their atmospheric analysis
-- they did not see N~\3 emission in their (lower-resolution)
spectrum, but N~\4 emission remains at \lam4058, which argues for an
earlier type.

The use of an O3 type to describe both Sk 183 and MPG 324 arises from
the high-quality of the spectra, in which the N~\4 emission is clearly
seen. These classifications are based on the spectroscopic framework
in this regime of massive stars, which have unambiguous criteria
linked to the relative intensities of the nitrogen lines (Walborn et
al.  2002, 2004; Sota et al. 2011).

\subsection{Stellar Radial Velocities}\label{rv}

Stellar radial velocities ($v_{\rm r}$) for Sk\,183 were determined
from Gaussian fits to the He~\2 \lam\lam4200, 4542 lines. From
measurements of each LR02 observation we found weighted means of
$v_{\rm r}$\,$=$\,162.5\,$\pm$\,1.2 and 162.4\,$\pm$\,0.8\,\kms\/ from
He~\2 \lam4200 and \lam4542, respectively. The quoted errors are
1-$\sigma$ uncertainties on the positions of the centroids of the
fits. A statistical variability analysis of the individual radial
velocity measurements does not detect significant differences,
suggesting a constant velocity from our spectroscopy. From fits to
the superimposed H$\alpha$ and [O~\3] \lam\lam4959, 5007 nebular
emission, the mean gas velocity is slightly offset with $v_{\rm
  gas}$\,$=$\,172.1\,$\pm$\,1.1\,\kms. This is approximately 5\,\kms\/
lower than the gas velocity measured at the same right ascension (but
$\sim$20$''$ further south) by Nigra et al. (2008).

\section{Physical properties of Sk\,183}

\subsection{Distance and Extinction}\label{reddening}

The distance to Sk\,183 is the largest uncertainty on its total luminosity.
Given the long-standing questions regarding the depth and structure of
the SMC, the `standard' distance modulus of $\sim$18.9\,mag (e.g.
Harries et al. 2003) might not be appropriate for the Wing. There is
mounting evidence for objects in the eastern part of the SMC to
suggest shorter distances, e.g., Howarth (1982), Mathewson et al.
(1986), Crowl et al. (2001), and Glatt et al. (2008). Indeed, in their
analysis of the {\em HST} imaging, Cignoni et al. (2009) advocate a
distance modulus to the young population in NGC\,602 of 18.7\,mag.

Sk\,183 is relatively isolated from bright companions so, to determine
the line-of-sight reddening, optical $UBVR$ photometry was taken from
Massey (2002). We also employed the $F814W$ magnitude from Schmalzl et
al. (2008) and near-infrared $JHK_{\rm s}$ magnitudes from the Two
Micron All Sky Survey (2MASS; Skrutskie et al.  2006), as shown in
Figure~\ref{fits} and summarised in Table~\ref{obsphot}.  Photometry
of Sk\,183 at longer wavelengths has become available recently from
the {\em Wide-field Infrared Survey Explorer} ({\em WISE}, Wright et
al. 2010); these data are discussed separately in Section~\ref{wisephot}.

\begin{center}
\begin{deluxetable}{lcl}
\tabletypesize{\footnotesize}
\tablewidth{0pc}
\tablecolumns{3}
\tablecaption{Photometry of Sk\,183\label{obsphot}}
\tablehead{\colhead{Band}  & \colhead{Magnitude} & \colhead{Ref.}}
\startdata
$U$ & $\phantom{^\dagger}$12.51\,$\pm$\,0.01$^\dagger$ & Massey (2002) \\
$B$    & $\phantom{^\dagger}$13.59\,$\pm$\,0.01$^\dagger$ & Massey (2002) \\
$V$    & 13.82\,$\pm$\,0.01 & Massey (2002) \\
$R$ & $\phantom{^\dagger}$13.89\,$\pm$\,0.02$^\dagger$ & Massey (2002) \\
$F814W$ & 14.128\,$\pm$\,0.003 & Schmalzl et al. (2008) \\
$J$ & 14.426\,$\pm$\,0.029 & 2MASS \\
$H$ & 14.606\,$\pm$\,0.063 & 2MASS \\
$K_{\rm s}$ & 14.618\,$\pm$\,0.091 & 2MASS \\
$W$1 (3.4\,$\mu$m) & 14.480\,$\pm$\,0.029 & {\em WISE} \\
$W$2 (4.6\,$\mu$m) & 14.255\,$\pm$\,0.041 & {\em WISE} \\
$W$3 (12\,$\mu$m) & 10.368\,$\pm$\,0.043 & {\em WISE} \\
$W$4 (22\,$\mu$m) & \ph5.955\,$\pm$\,0.031 & {\em WISE}
\enddata
\tablecomments{$^\dagger$ Calculated in quadrature from the errors
  quoted on the colours by Massey (2002).}
\end{deluxetable}
\end{center}

\vspace{-0.35in}From fits of the adopted model atmosphere (Section~\ref{analysis}) to
the spectral energy distribution of the {\it IUE} spectroscopy and
multi-band photometry (Figure~\ref{fits}) we derived
$E(B-V)$\,$=$\,0.09\,$\pm$\,0.01\,mag. This included contributions
from both internal SMC and foreground Galactic reddening, taking the
lower bound of $E(B-V)$\,$=$\,0.04\,mag from Bessell (1991) for the
latter. In calculating the total visual extinction we adopted ratios
of total-to-selective extinction of 3.1 (Galactic; Howarth, 1983) and
2.7 (SMC; Bouchet et al. 1985), leading to $A_V$\,$=$\,0.26\,mag and
an absolute magnitude of $M_{V}$\,$=$\,$-$5.14 or $-$5.34\,mag (for
distance moduli of 18.7 and 18.9\,mag, respectively). Such values are at
the faint end of those found for early-type dwarfs (e.g. Walborn et
al. 2002), and are comparable to others classified as Vz (see, e.g.,
Figure 4 from Walborn 2009), notwithstanding a potential contribution
from a companion.

\subsection{Atmospheric Analysis}\label{analysis}

\subsubsection{PoWR Models}
Physical properties for Sk\,183 were estimated from comparisons of the
FLAMES data with synthetic spectra calculated with the PoWR code (e.g.
Hamann \& Gr\"{a}fener, 2003; 2004). PoWR treats the wind and the
photosphere self-consistently and has been used to analyse the wind
characteristics of O-type stars (Oskinova et al. 2006, 2007),
helium-rich subdwarf O-type stars (Jeffery \& Hamann, 2010), and
B-type main-sequence stars (Oskinova et al. 2011)\footnote{Analogous
  to, e.g., the differences discussed by Rivero Gonz\'{a}lez et al.
  (2011) between results from {\sc cmfgen} (Hillier \& Miller, 1998)
  and {\sc fastwind} (Puls et al.  2005), we note there may be small
  systematic differences if the parameters determined from PoWR were
  compared to those obtained with other codes.}.

In brief, the PoWR code solves the radiative transfer equation for a
spherically-expanding atmosphere and the statistical equilibrium
equations simultaneously, while also accounting for energy
conservation and allowing deviations from local thermodynamic
equilibrium (i.e.  non-LTE). A PoWR model is defined by its effective
temperature ($T_{\rm eff}$), effective surface gravity ($\log g_{\rm
  eff}$), luminosity ($L$), mass-loss rate ($\dot{M}$), terminal
velocity of the wind ($v_\infty$), and chemical composition. A
standard $\beta$-velocity law ($\beta$\,$=$\,1) for the wind
acceleration was assumed in all calculations; the density follows from
the velocity via the mass-continuity equation. The wind domain is
smoothly connected to the photosphere, at which point the hydrostatic
equilibrium is approached asymptotically.

The models included transitions of hydrogen, helium, carbon, oxygen,
nitrogen, and silicon. Iron and iron-group elements were included
using the concept of superlevels, as described by Gr\"{a}fener et al.
(2002); this is important because of the blanketing effect of the
summed opacities on the atmospheric structure.

\subsubsection{`Single-star' Analysis}\label{ss_analysis}
We first assumed that the observed spectrum originates from a single
star.~Stellar parameters were determined iteratively, with the
bolometric luminosity calculated from the absolute magnitude
($-$5.14\,mag) from Section~\ref{reddening}. The model spectra were
convolved with a rotational-broadening function to take into account
an apparent projected rotational velocity of $v{\rm
  sin}i$\,$=$\,60\,$\pm$\,20\,\kms, determined from the metal
  lines (N~\3, N~\4, N~\5, and Si~\4).  The primary temperature
diagnostic available is the nitrogen ionization balance, with
additional constraints given by the Si~\4 emission and the He~\1
\lam4471 absorption.

\begin{figure*}
\begin{center}
\includegraphics[width=17cm]{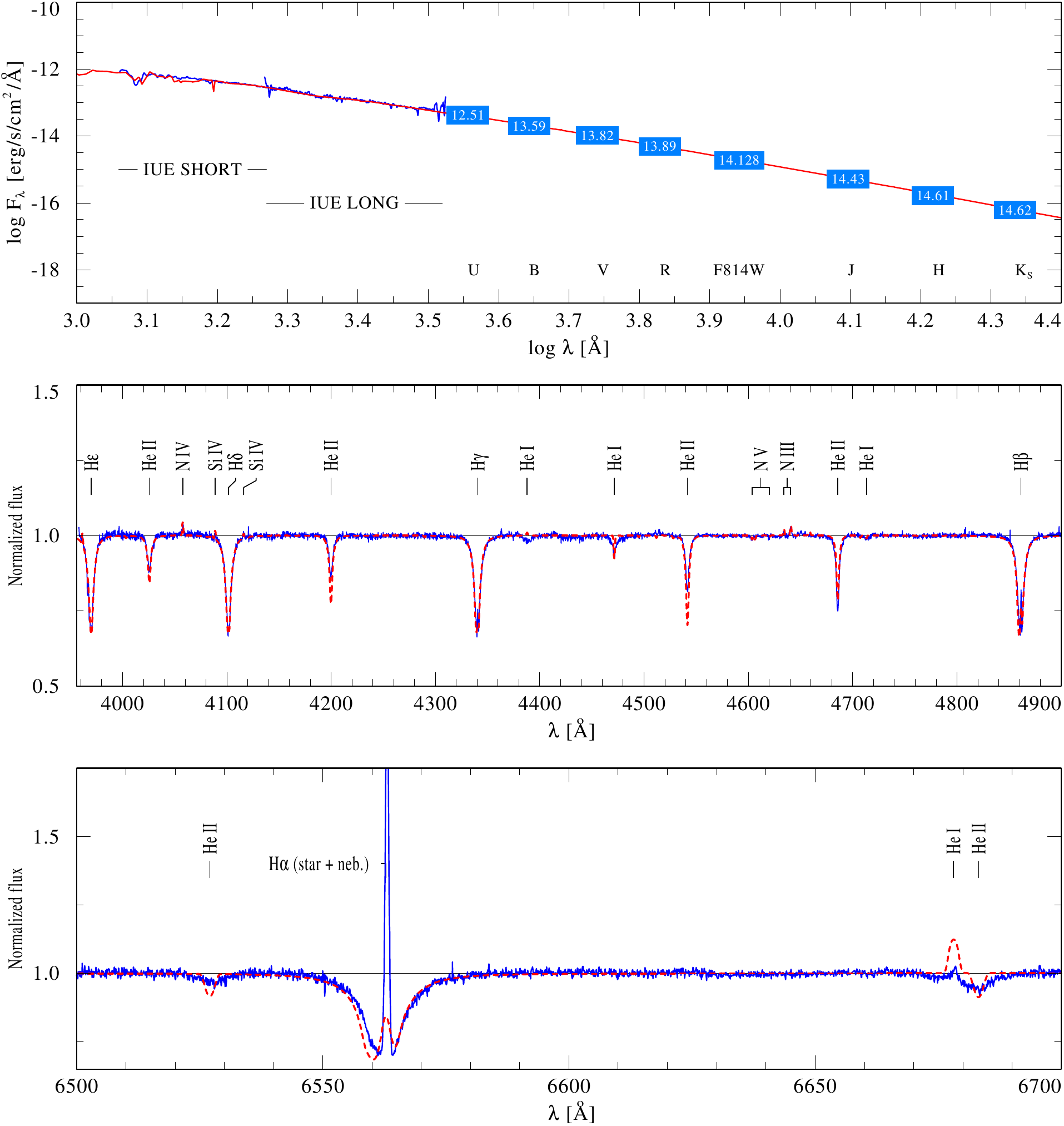}
\caption{{\it Upper panel:} model fit to the spectral energy
  distribution of Sk\,183, combining {\it IUE} spectroscopy and
  multi-band photometry as indicated. The total line-of-sight
  reddening, taking into account Galactic and SMC contributions, is
  $E(B-V)$\,$=$\,0.09\,mag. {\it Middle and lower panels:} comparison
  of the optical FLAMES spectrum of Sk\,183 (blue solid line) with the
  adopted `single star' PoWR model (red dashed line).
}\label{fits}
\end{center}
\end{figure*}

From comparisons of model spectra with the observations, we adopted a
nitrogen-rich model with $T_{\rm eff}$\,$=$\,46$\pm$2\,kK and
$\log{g_{\rm eff}}$\,$=$\,4.0\,$\pm$\,0.1, as shown in
Figure~\ref{fits}.  The stellar radius, at a Rosseland optical depth
of 20, is $R_{\ast}$\,$=$\,9.7\,$R_{\odot}$. The He~\2 \lam\lam4200,
4542 lines were somewhat over-predicted by this model. While a cooler model
($T_{\rm eff}$\,$=$\,42\,kK) still provided a reasonable fit to the
Si~\4 emission, it led to significant over-prediction of the He~\1
absorption with little improvement for the He~\2 lines. A simultaneous
fit to the (notably broader) weaker He~\1 lines was not possible
with a model of a single star.

The adopted 46\,kK model fits the observed N~\3, N~\4, and N~\5
features simultaneously with an abundance of
12+$\log$([N/H])\,$=$\,7.73 ($\pm$20\%). Compared with the Solar value
of 7.83 (Asplund et al. 2009) this suggests significant enrichment
compared to the baseline SMC value found from H~\2 regions
($\sim$6.55, Russell \& Dopita, 1990). The uncertainty quoted on the
$T_{\rm eff}$ estimate ($\pm$2\,kK) reflects the fitting of the
stellar parameters.  With more modest nitrogen abundances, e.g., a
factor of two lower (comparable to that obtained for NGC\,346 MPG\,355
by Walborn et al.  2004), the N~\4 and N~\5 features could be fit with
a hotter temperature (of a few kK), but a simultaneous fit of the N~\3
lines was not possible.

The mass-loss rate appears to be low, as might be expected given the
reduced metallicity of the SMC (e.g. Bouret et al.  2003).~We
calculated a grid of models where all parameters were fixed, but with
a range of
10$^{-9}$\,$\le$\,$\dot{M}$\,$\le$\,10$^{-5}$\,$M_\odot$\,yr$^{-1}$.~From
comparison of the models with the observed spectra,
$\dot{M}$\,$=$\,10$^{-7}$\,$M_\odot$\,yr$^{-1}$ was the best
compromise to fit the observed UV (N~\5) and optical (H$\alpha$ and
He~\2 \lam4686) profiles. For example, if
$\dot{M}$\,$>$\,10$^{-7}$\,$M_\odot$\,yr$^{-1}$ the predicted He~\2
\lam4686 absorption was weaker than observed, suggesting this as an
upper limit. A more precise estimate of $\dot{M}$ would
require high-quality UV and/or $K$-band spectroscopy.

To estimate $v_\infty$ we employed the empirical relation from Prinja
(1994) between $v_\infty$ and the wavelength difference ($\Delta$\lam)
between the observed peak emission and minimum absorption of the N~\5
or C~\4 lines in low-resolution {\it IUE} spectra. Using Prinja's
equation\,(1) for Sk\,183, $\Delta$\lam(N~\5)\,$\approx$\,12\,\AA, giving
$v_\infty$\,$\approx$\,2150\,\kms\/ ($\pm$\,$\sim$300\,\kms); this
value was adopted in our models.

The degree of clumping in the wind was poorly constrained by the
available data because of the low mass-loss rate. The final model
was therefore calculated without inclusion for the effects of clumping, 
but note that the exact values of $v_\infty$ and the clumping factor did
not have a significant impact on the other physical parameters. In
particular, the N~\4 emission in our PoWR model does not arise from
the wind, so was not affected by the adopted clumping factor.

When correcting the effective gravity for the contribution of radiation
pressure we found $\log g$\,$=$\,4.11, yielding a spectroscopic mass
estimate of $M_{\rm spec}$\,$=$\,46$^{+9}_{-8}$ $M_\odot$. This is in
good agreement with the evolutionary mass obtained from comparisons
with the theoretical evolutionary tracks for the SMC from Brott et al.
(2011). For \vsini\,(init.)\,$<$\,350\,\kms, the $T_{\rm eff}$ and $L$
determined for Sk\,183 lies roughly equidistant between the 40 and
50\,$M_\odot$ tracks, suggesting an evolutionary mass for Sk\,183 of
45\,$\pm$\,5$M_\odot$, with a corresponding age of $\sim$2\,Myr.  In
the LMC grid from Brott et al. there are also models for an initial
mass of 45\,$M_\odot$. Scaling the moderate rotation models
(\vsini\,(init.)\,$<$\,350\,\kms) to the hotter temperatures obtained
for stars in the SMC, we find reasonable agreement with the parameters
obtained for Sk\,183 (Dr~I.~Brott, private communication).

The parameters determined from the single-star analysis of Sk\,183 are
summarised in Table~\ref{params}. For completeness, in the third
column we provide the equivalent results assuming the canonical
distance modulus to the SMC of 18.9\,mag.

\begin{center}
\begin{deluxetable*}{lllll}
\tablecaption{Derived Properties of Sk\,183 \label{params}}
\tablewidth{0pc}
\tablecolumns{4}
\tablehead{ & \multicolumn{2}{c}{Single-star Model} & \multicolumn{2}{c}{Composite Model} \\
& \colhead{DM\,$=$\,18.7\,mag} & \colhead{DM\,$=$\,18.9\,mag} & \colhead{Pri.} & \colhead{Sec.}}
\startdata
$M_{V}$ & $-$5.14\,mag & $-$5.34\,mag & \multicolumn{2}{c}{$-$5.14\,mag} \\
$\log L/L_{\odot}$ & 5.58\,$\pm$\,0.02 & 5.66\,$\pm$\,0.02 & 5.57 & 4.1 \\
$T_{\rm eff}$ & \multicolumn{2}{c}{46\,$\pm$\,2\,kK} & 47.5\,kK & 21\,kK \\
$R_{\ast}$ & 9.7\,$R_{\odot}$ & 10.7\,$R_{\odot}$ & 9.6\,$R_\odot$ & 8.5\,$R_\odot$ \\
$\log$ $g_{\rm eff}$ & 4.0\,$\pm$\,0.1 & 4.0\,$\pm$\,0.1 & 4.0 & 3.5 \\ 
$\log$ $g$ & 4.11$^{-0.07}_{+0.08}$ & 4.11$^{-0.07}_{+0.08}$ & 4.11 & 3.5 \\
$M_{\rm spec}$ & 46$^{+9}_{-8}$\,$M_\odot$& 54$^{+11}_{-7}$\,$M_\odot$& 40$^{+8}_{-6}$\,$M_\odot$ & 8.7\,$M_\odot$ \\
$\dot{M}$ & \multicolumn{2}{c}{10$^{-7}$ $M_{\odot}$ yr$^{-1}$} & 10$^{-7}$ $M_{\odot}$ yr$^{-1}$ & ... \\
\vsini & \multicolumn{2}{c}{60\,$\pm$\,20\,\kms} & 60\,\kms & 250\,\kms \\
$v_{\infty}$ & \multicolumn{2}{c}{2150\,$\pm$\,300\,\kms} & 2150\,\kms & ... \\
$\log$ $Q_{\rm 0}$ & 49.37\,s$^{-1}$ & 49.45\,s$^{-1}$ & 49.41\,s$^{-1}$ & ... \\ 
12+$\log$([N/H]) & \multicolumn{2}{c}{7.73 ($\pm$20\%)} & 7.73 & ... 
\enddata
\tablecomments{The quoted errors on the single-star models are from 
fitting uncertainties. We do not quote uncertainties on the composite
model as it is non-unique fit to the data. }
\end{deluxetable*}
\end{center}

\vspace{-0.3in}
\subsubsection{Composite Model}

We also investigated a scenario in which the observed spectrum of
Sk\,183 is comprised of a hot O-type primary, combined with a cooler,
lower luminosity, secondary component.  Composite spectra were
synthesized by scaling the continuum fluxes of the two components by
their relative luminosities, such that the combined model reproduces
the observed spectral energy distribution of Sk\,183 (assuming a
distance modulus of 18.7\,mag). Given the large luminosity of the O3
component, the total contribution from, e.g., a B-type companion, is
relatively small (indeed, it is within the uncertainty of the
single-star fit). 

\begin{figure*}
\begin{center}
\includegraphics[width=17cm]{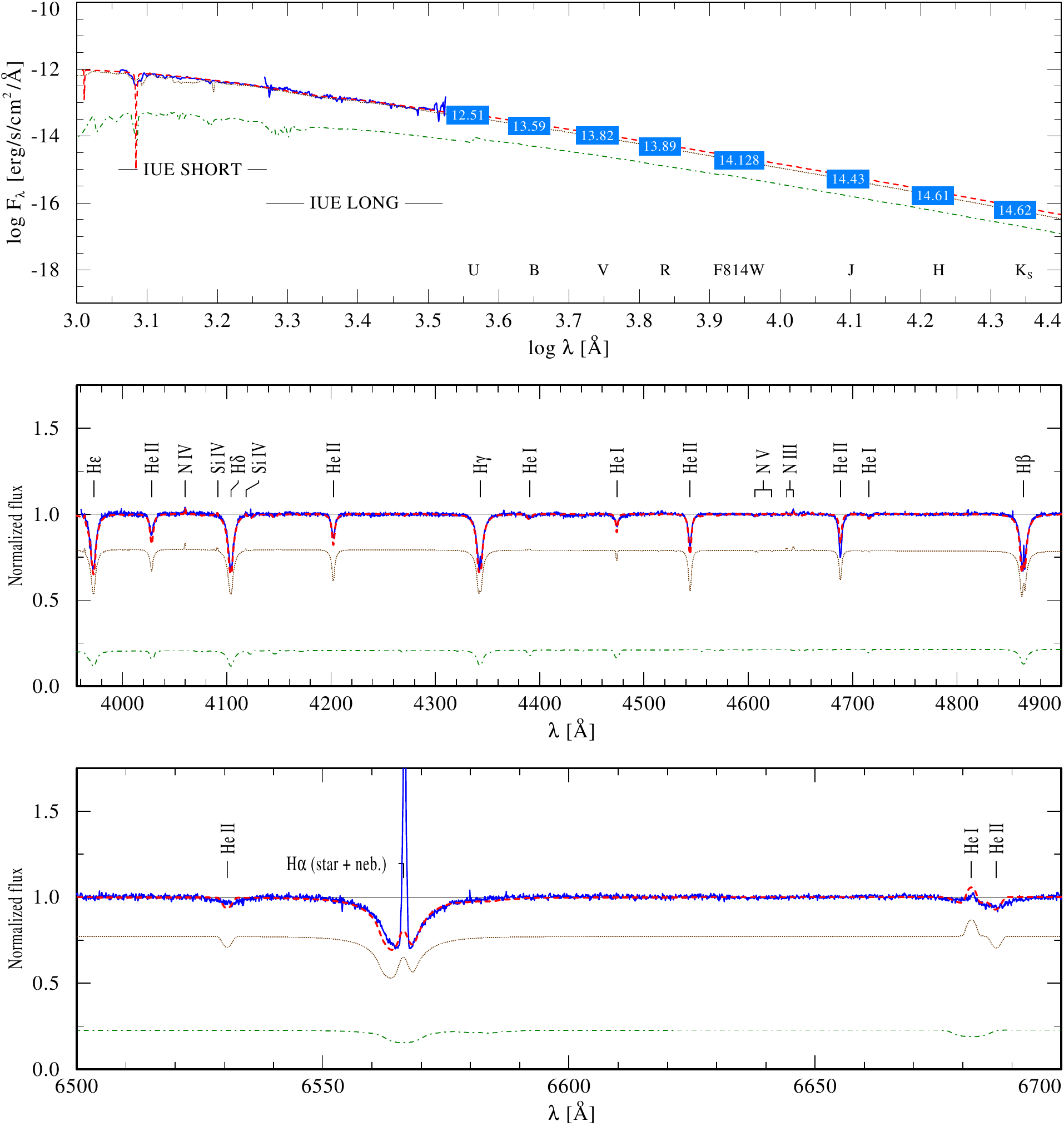}
\caption{As Figure~\ref{fits} but for a composite synthetic spectrum
  (dashed, red line), comprised of a hot O-type model ($T_{\rm
    eff}$\,$=$\,47.5\,kK; dotted, brown line) and an early B-type
  component ($T_{\rm eff}$\,$=$\,21\,kK; dash-dotted, green line).
  The latter two are plotted with their fluxes scaled by their
  relative intensities.}\label{fits_bin}
\end{center}
\end{figure*}

Adopting the same radial velocity for the secondary spectrum, a
reasonable fit was obtained with a composite of a slightly hotter,
early O-type model ($T_{\rm eff}$\,$=$\,47.5\,kK) with that of an
early B-type giant (with $\log L/L_{\odot}$\,$=$\,4.1, $T_{\rm
  eff}$\,$=$\,21\,kK, and $\log$ $g_{\rm eff}$\,$=$\,3.5), as shown in
Figure~\ref{fits_bin}.  A large (projected) rotational velocity of
\vsini\,$=$\,250\,\kms\/ was necessary for the secondary component to
reproduce the broad He~\1 lines.

While there is only tentative evidence of a secondary companion, we
include this analysis to explore its potential consequences on the
combined spectrum. Indeed, if a genuine binary, it would be surprising
(in an evolutionary sense) to find a B-type giant with a much younger
O3-type dwarf.  However, this is not a unique solution for
the secondary, as hotter models for the secondary (e.g. a late O-type
star) also give reasonable fits. 

\begin{figure}
\begin{center}
\includegraphics[width=8.25cm]{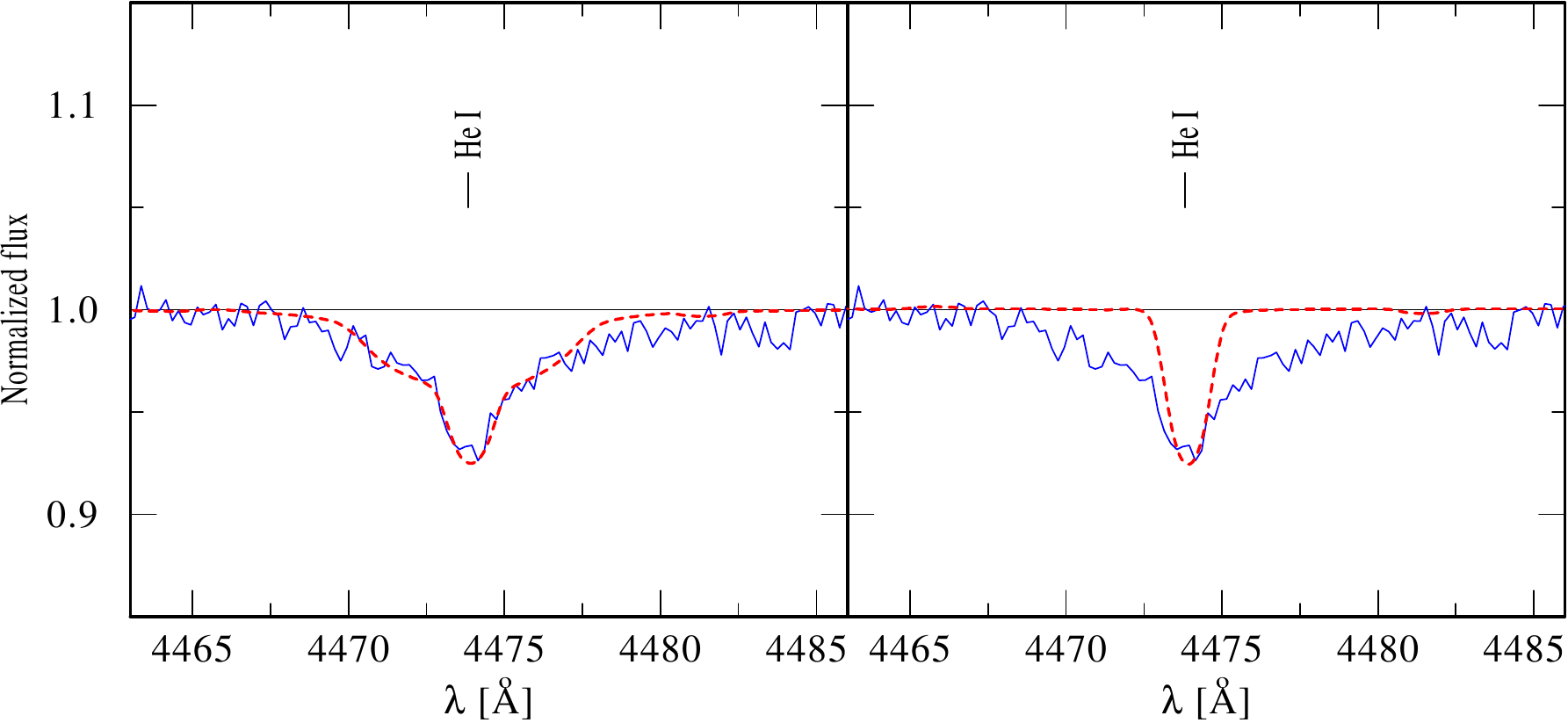}
\caption{{\it Left-hand panel:} composite model (red; $T_{\rm
    eff}$\,$=$\,47.5 and 21\,kK for the two components) of the
  observed He~\1 \lam4471 profile. {\it Right-hand panel:}
  `single-star' fit ($T_{\rm eff}$\,$=$\,46\,kK).}\label{4471}
\end{center}
\end{figure}

The putative composite model has some advantages compared to the
single-star fit. In addition to the fits to the broad He~\1 lines, the
reproduction of the He~\2 \lam\lam4200, 4542 lines is improved due to
dilution of the continuum level by the second component. Moreover, a
more consistent fit is achieved for the He~\1 \lam4471 line -- we were
unable to reproduce the broad wings and narrow core of this line
simultaneously in the single-star model (see Figure~\ref{4471}). The
parameters of the primary and secondary components in the composite
model are summarised in Table~\ref{params}.

\section{Discussion}

The discovery of such an early-type star in the SMC, let alone in the
Wing, is noteworthy. To place its rarity in context, there are
published classifications for over 5000 luminous stars in the SMC (e.g. see
compendium by Bonanos et al. 2010) but only two with secure spectral
types of earlier than O4, namely NGC\,346 MPG~355 and
AzV\,435\footnote{MPG~355 was classified as O2 III(f*) by Walborn et
  al. (2002) and ON2 III(f*) by Walborn et al. (2004); AzV\,435 was
  classified as O3 V((f*)) by Massey et al. (2005). Note that the
  spectral type listed by Bonanos et al.  for AzV~14 is O3-4~V
  (Garmany et al. 1987), but this was revised to O5~V by Massey et al.
  (2004).}.  Thus, including the revised classification of MPG~324,
the new spectroscopy of Sk\,183 brings the total of O2 and O3 stars
known in the SMC to only four objects.

These stars provide a glimpse of some of the most massive O-type stars
in a very metal-poor environment, into a domain of significantly
diminished mass-loss rates (e.g. Bouret et al. 2003). Indeed, in
their discussions of chemically-homogeneous evolution in the context
of gamma-ray bursts, Yoon et al. (2006) noted NGC\,346 as the only
obvious cluster in which to examine the evolution of the most massive
stars in a truly metal-poor regime; Sk\,183 provides an excellent
additional case for further study.  In particular, spectroscopic
monitoring would be helpful to unveil its true nature.

\subsection{Ionisation of the N90 Nebula}

The {\em Spitzer} observations revealed a population of young stellar
objects along the inner edge of the N90 nebula, with Carlson et al.
(2007) concluding that star formation started in the central cluster
$\sim$4\,Myr ago and has propagated outwards since then. The peak star
formation rate in the region has occurred over the past 2.5\,Myr
(Cignoni et al. 2009), during which time it has been approximately
constant (Carlson et al. 2011). The evolutionary age obtained for
Sk\,183 ($\sim$2\,Myr, Section~\ref{ss_analysis}), is consistent with
it forming in this recent phase of star formation.  Meanwhile, the gas
in the interior of the H~\2 region is nearly quiescent and does not
show violent motions (Nigra et al. 2008), leading Carlson et al.
(2011) to conclude that photoionisation is the primary trigger of the
current star formation.

N90 has been observed from the ground and with the {\em HST} in
narrow-band filters, thus isolating the H$\alpha$ and [N~\2] emission.
Using the ACS images obtained in program GO-10284 (P.I.: A. Nota) we
integrated the flux from the high surface-brightness component of the
nebula finding
$\sim$2\,$\times$\,$10^{-11}$\,erg\,s$^{-1}$\,cm$^{-2}$, after
corrections for the foreground extinction and the weak [N~\2]
emission.  The former was taken as $A_R$\,$=$\,0.16\,mag
from the NASA/IPAC Infrared Science Archive Extinction
Calculator (which uses the results from Schlegel et al. 1998), 
while a contribution to the [N~\2] emission of 4\% was adopted
(e.g., Pe\~{n}a-Guerrero et al. 2012).
However, the ACS image covers only the core part of N90, which extends
over a larger area of 8$'$\,$\times$\,6$'$.  Thus, integrating over
the entire extent of N90 from the Magellanic Cloud Emission Line
Survey (MCELS; Smith et al.  1999) we obtained
$f_{0}$(H$\alpha$)\,$\approx$\,8\,$\times$\,$10^{-11}$\,erg\,s$^{-1}$\,cm$^{-2}$
(cf.
$f_{0}$(H$\alpha$)\,$=$\,10.8\,$\pm$1.1\,$\times$\,$10^{-11}$\,erg\,s$^{-1}$\,cm$^{-2}$
from Kennicutt \& Hodge, 1986). The flux from the MCELS image
corresponds to a total luminosity of
$\ge$\,3.2\,$\pm$\,0.3\,$\times$10$^{37}$\,erg\,s$^{-1}$ (with the
uncertainty indicative of the two distance moduli, i.e. 18.7 and
18.9\,mag).

Converting this $L$(H$\alpha$) to a Lyman continuum luminosity for a
standard nebular case yields an ionization rate of $Q_{0} \approx$ 2
$\times 10^{49}$~s$^{-1}$. This will be a lower limit as some
ionization surfaces within N90 are hidden from optical view by dark
clouds and we have not accounted for leakage. Aside from Sk\,183, the
next earliest objects known in NGC\,602a are stars \#5 and \#2
(Hutchings et al. 1991, classified as O8 and O9, respectively).  The
ionising fluxes from late O-type stars are over an order of magnitude
lower (see, e.g., Smith et al. 2002), such that Sk\,183 appears to be
the dominant source of ionisation.

\subsection{Mid-infrared Excess}\label{wisephot}

\begin{figure}
\begin{center}
\includegraphics[width=8.5cm]{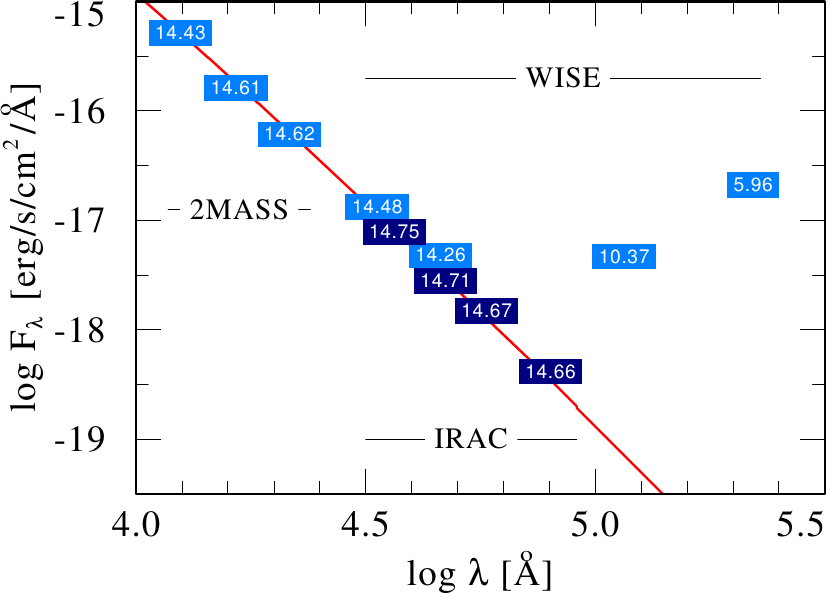}
\caption{2MASS, {\em WISE}, and {\em Spitzer}-IRAC photometry
  toward Sk\,183, compared with the best fitting single-star model
  from Figure~\ref{fits}.  The longer-wavelength {\em WISE} observations
  ($W$3\,$=$\,12\,$\mu$m and $W$4\,$=$\,22\,$\mu$m) reveal an apparent
  mid-IR excess.  Inspection of the {\em Spitzer} 24\,$\mu$m image
  suggests that this excess arises from an extended source slightly
  offset from Sk\,183 (see Figure~\ref{spitzer}).}\label{wise}
\end{center}
\end{figure}

The spectral energy distribution of Sk\,183 is well matched in the
near-IR by both the single-star and composite models
(Figures~\ref{fits} and \ref{fits_bin}).  However, inspection of the
new {\em WISE} photometry (Table~\ref{obsphot}) reveals an interesting
upturn in the flux distribution, as shown in Figure~\ref{wise}.
Observations with the short-wavelength {\em WISE} bands ($W$1 and $W$2,
corresponding to 3.6 and 4.6\,$\mu$m, respectively) are in good
agreement with the predicted flux from our PoWR models.  But at the
longer wavelengths of the $W$3 and $W$4 bands (12 and 22\,$\mu$m), there
is a clear mid-IR excess compared to the predicted distribution. A
peak of the mid-IR flux distribution at 22\,$\mu$m would argue for a
blackbody temperature of order 130\,K, or even cooler if the flux
peaks at longer wavelengths. 

Similar mid-IR excesses for otherwise normal OB-type stars have been
seen in the SMC (Bolatto et al. 2007; Ita et al. 2010; Bonanos et al.
2010) and LMC (Evans et al.  2011). To investigate the mid-IR excess
in Sk\,183 further, we obtained images of the region from the {\em
  Spitzer} SAGE-SMC Survey (Gordon et al. 2011).

\begin{figure*}
\begin{center}
\includegraphics[width=17cm]{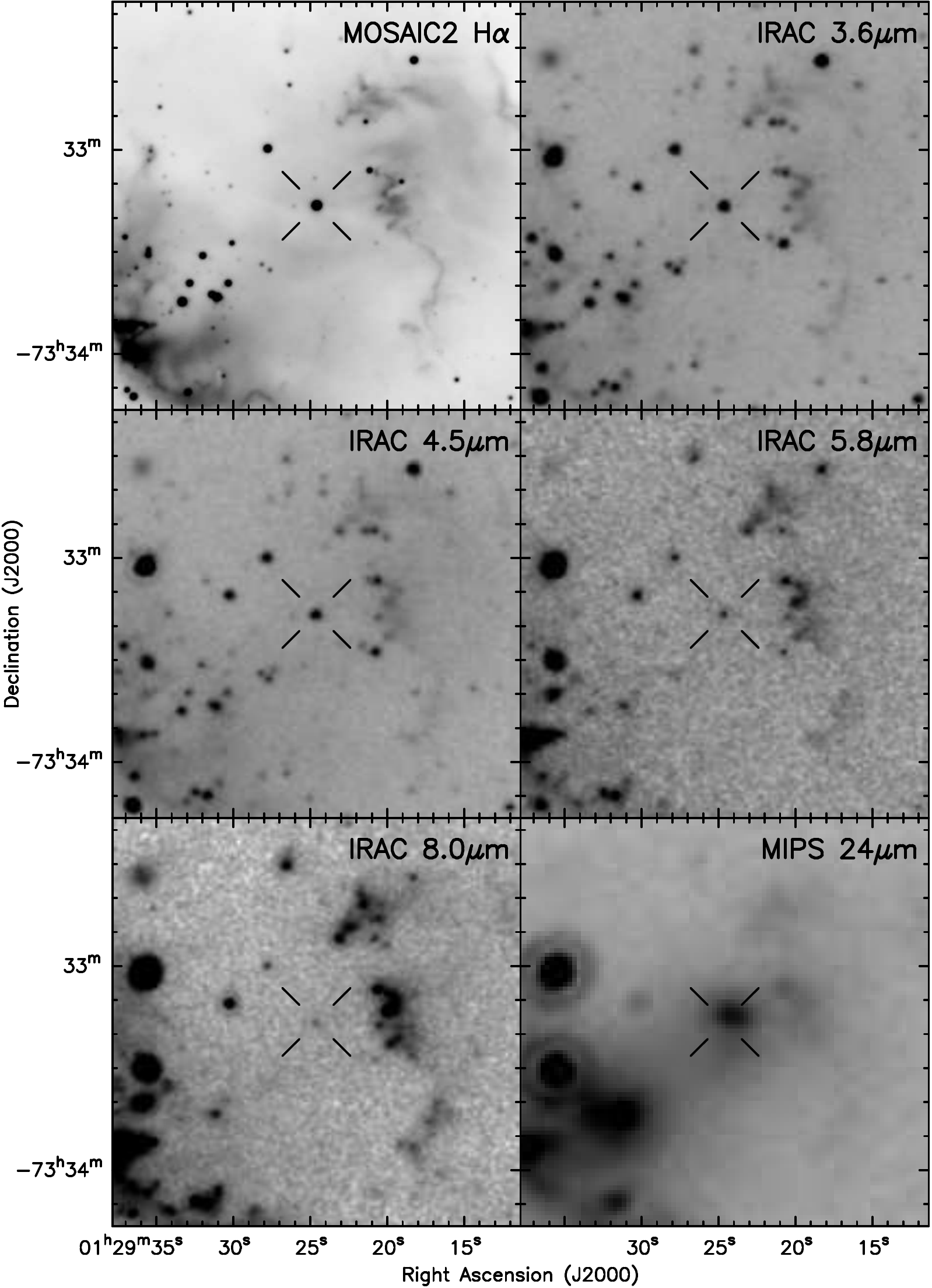}
\caption{Mosaic\,II (H$\alpha$), {\em Spitzer} IRAC, and MIPS
  (2\arcmin$\times$2\arcmin) images of Sk\,183 (marked by the open
  cross) and the surrounding nebula.  In the MIPS 24~$\mu$m image the
  mid-IR excess emission suggested by the {\em WISE} photometry (see
  Section~\ref{wisephot}) appears to arise from an extended source
  slightly offset from the location of Sk\,183.}\label{spitzer}
\end{center}
\end{figure*}

The InfraRed Array Camera (IRAC, 3.6, 4.5, 5.8, 8\,$\mu$m) and
Multiband Imaging Photometer for {\em Spitzer} (MIPS, 24\,$\mu$m)
images of Sk\,183 are shown in Figure~\ref{spitzer}, together with an
H$\alpha$ image obtained with the Mosaic~II camera on the 4-m Blanco
telescope at the Cerro Tololo Inter-American Observatory.  From
photometric extractions of the IRAC images we found:
[3.6]\,$=$\,14.75, [4.5]\,$=$\,14.71, [5.8]\,$=$\,14.67, and
[8.0]\,$=$\,14.66\,mag.  However, the peak of the 24\,$\mu$m flux is
offset slightly and appears extended (cf. the Airy rings
seen around the point sources to the east), suggesting it as nearby
dust emission, or unresolved dusty shocks (e.g. those seen in the
Carina nebula, Smith et al. 2010), but not directly associated with
Sk\,183.

\subsection{High-mass Star Formation in NGC\,602}

In the standard model of star formation, the initial mass function
(IMF) is purely statistical in nature and the probability of very
massive stars being formed, while low, is independent of the mass of
the star-forming region (e.g. Parker \& Goodwin, 2007). In the
integrated galactic IMF model (IGIMF, e.g., Weidner \& Kroupa 2006)
this probability increases with the mass of the star-forming region,
with the prediction that the most massive stars should be found in the
largest star-forming complexes.

NGC\,602a is a relatively low mass cluster, with a stellar mass of
$\sim$2000~$M_{\odot}$ derived from the combination of the optical
(Cignoni et al. 2009) and infrared imaging (Carlson et al.  2011). In
this context, the IGIMF predictions from Weidner et al. (2011) suggest
a maximum stellar mass of $\sim$50 to 60~$M_{\odot}$, roughly
consistent with the estimated mass of the primary of Sk\,183.
Intriguingly though, Sk\,183 is located approximately 45$''$ to the
north-west of NGC\,602a (corresponding to a projected distance of 
12--13\,pc, depending on the adopted distance modulus). This suggests
the star has formed in {\it relative} isolation compared to the main
body of the cluster (with no obviously massive cluster visible within
3\,pc, e.g., Bressert et al. 2012), or that it is a slow runaway from
NGC\,602a, as alluded to by Gouliermis et al. (2012). Analysis of the
line-of-sight velocities of other stars in N90 (and the gas profiles
super-imposed on their spectra) is underway and should provide a more
complete picture of the dynamics of this region.

While the single case of NGC\,602 can not distinguish between the two
IMF models, the SMC is rich in small, star-forming regions with masses
in the range of 1000--2000\,$M_{\odot}$. A more complete census of
O-type stars in SMC clusters would provide an interesting test of the
theories of the high-mass IMF, complementing recent work investigating
`isolated' O-type stars in the SMC by Lamb et al. (2010) and Selier et
al. (2011).

\section*{Acknowledgements}
Based on observations obtained in ESO program 086.D-0167. We are
grateful to Nolan Walborn and Ines Brott for useful discussions, and
to the referees for their constructive comments. We acknowledge DFG
Grant OS~292/3-1 which supported a meeting of the authors, and LMO
acknowledges DLR Grant 50\,OR\,1101. JSG thanks the University of
Wisconsin-Madison Graduate School for partial support of this work.
JSG, YHC, and RAG are supported by the NASA Grant SAO~GO0-11025X.  VHB
acknowledges support from the Scottish Universities Physics Alliance
(SUPA) and from the Natural Science and Engineering Research Council
of Canada (NSERC).  This research has made use of the NASA/IPAC
Extragalactic Database (NED) which is operated by the Jet Propulsion
Laboratory, California Institute of Technology, under contract with
NASA. This publication also makes use of data products from the
{\em Wide-field Infrared Survey Explorer}, which is a joint project of the
University of California, Los Angeles, and the Jet Propulsion
Laboratory/California Institute of Technology, funded by NASA.

\end{document}